\documentclass[prl,twocolumn,superscriptaddress, amsmath,amssymb,]{revtex4-1}
\pdfoutput=1
\usepackage{graphicx}
\usepackage{dcolumn}
\usepackage{bm}
\usepackage{siunitx}

\begin{document}

\title{Cation- and lattice-site-selective magnetic depth profiles of ultrathin $\mathrm{Fe_3O_4}$(001) films}
\author{\underline{Tobias Pohlmann}}
\email{tobias.pohlmann@desy.de}
\affiliation{Department of Physics, Osnabr\"uck University, Barbarastr. 7, 49076 Osnabr\"uck, Germany}
\affiliation{DESY Photon Science, Notkestr. 85, 22607 Hamburg, Germany}
\author{Timo Kuschel}
\affiliation{Center for Spinelectronic Materials and Devices, Department of Physics, Bielefeld University, Universit\"atsstr. 25, 33615 Bielefeld, Germany}
\author{Jari Rodewald}
\affiliation{Department of Physics, Osnabr\"uck University, Barbarastr. 7, 49076 Osnabr\"uck, Germany}
\author{Jannis Thien}
\affiliation{Department of Physics, Osnabr\"uck University, Barbarastr. 7, 49076 Osnabr\"uck, Germany}
\author{Kevin Ruwisch}
\affiliation{Department of Physics, Osnabr\"uck University, Barbarastr. 7, 49076 Osnabr\"uck, Germany}
\author{Florian Bertram}
\affiliation{DESY Photon Science, Notkestr. 85, 22607 Hamburg, Germany}
\author{Eugen Weschke}
\affiliation{Helmholtz-Zentrum Berlin f\"ur Materialien und Energie, Wilhelm-Conrad-R\"ontgen-Campus BESSY II, Albert-Einstein-Strasse 15, 12489 Berlin, Germany}
\author{Padraic Shafer}
\affiliation{Advanced Light Source, Lawrence Berkeley National Laboratory, 6 Cyclotron Rd, Berkeley, CA, 94720, USA}
\author{Joachim Wollschl\"ager}
\email{jwollsch@uni-osnabrueck.de}
\affiliation{Department of Physics, Osnabr\"uck University, Barbarastr. 7, 49076 Osnabr\"uck, Germany}
\author{Karsten K\"upper}
\email{kkuepper@uni-osnabrueck.de}
\affiliation{Department of Physics, Osnabr\"uck University, Barbarastr. 7, 49076 Osnabr\"uck, Germany}

\date{\today}
             
\begin{abstract}
A detailed understanding of ultrathin film surface properties is crucial for the proper interpretation of spectroscopic, catalytic and spin-transport data. We present x-ray magnetic circular dichroism (XMCD) and x-ray resonant magnetic reflectivity (XRMR) measurements on ultrathin $\mathrm{Fe_3O_4}$ films to obtain magnetic depth profiles for the three resonant energies corresponding to the different cation species $\mathrm{Fe^{2+}_{oct}}$, $\mathrm{Fe^{3+}_{tet}}$ and $\mathrm{Fe^{3+}_{oct}}$ located on octahedral and tetrahedral sites of the inverse spinel structure of $\mathrm{Fe_3O_4}$. By analyzing the XMCD spectrum of $\mathrm{Fe_3O_4}$ using multiplet calculations, the resonance energy of each cation species can be isolated. Performing XRMR on these three resonant energies yields magnetic depth profiles that correspond each to one specific cation species. The depth profiles of both kinds of $\mathrm{Fe^{3+}}$ cations reveal a $\mathrm{3.9 \pm 1\,\r{A}}$-thick surface layer of enhanced magnetization, which is likely due to an excess of these ions at the expense of the $\mathrm{Fe^{2+}_{oct}}$ species in the surface region. The magnetically enhanced $\mathrm{Fe^{3+}_{tet}}$ layer is additionally shifted about $\mathrm{3\pm 1.5\,\r{A}}$ farther from the surface than the $\mathrm{Fe^{3+}_{oct}}$ layer. 
\end{abstract}
\pacs{Valid PACS appear here}
\maketitle


\textit{Introduction.}$-$Magnetite, $\mathrm{Fe_3O_4}$, is one of the most frequently investigated transition-metal oxides, since it is a key material in spintronics \cite{Moussy13}, spin caloritronics \cite{Ramos13} and material chemistry \cite{Lin05}. $\mathrm{Fe_3O_4}$ thin films were considered highly suitable as electrode material for magnetic tunnel junctions \cite{Kado08,Marnitz15} due to their predicted half-metallic behavior with $100\%$ spin polarization \cite{Zhang91}. However, the promise was never quite met, with modest tunnel magnetoresistance ratios ranging from -26\% to 18\% \cite{Hu02,Kado08,Marnitz15}. In order to test its half-metallicity, spin-resolved x-ray photoelectron spectroscopy on $\mathrm{Fe_3O_4}$(111) films found a spin-polarization of about $80\%$ \cite{Dedkov02}, while on $\mathrm{Fe_3O_4}$(001) the same technique yielded polarizations of only $\approx 40\%-70\%$ \cite{Tobin07,Fonin08,Wang13}. 

Both the reduced tunnel magnetoresistance and the deviations from $100\%$ spin-polarization in spin-resolved XPS were argued to emerge from interface and surface effects, respectively \cite{Kado08,Marnitz15,Dedkov02,Fonin08,Wang13}. For $\mathrm{Fe_3O_4}$(111) films deposited on semiconducting ZnO(0001), lattice-site-selective depth profiles obtained by x-ray resonant magnetic reflectivity (XRMR) and electron energy loss spectroscopy did not find a notable surface modification apart from a $\mathrm{Fe_{oct}}$ termination \cite{Bruck12}. Reduction of the spin-polarization measured at the $\mathrm{Fe_3O_4}$(001) surface was typically considered to originate from a surface reconstruction, the existence of which has long been known but only recently has been resolved as a subsurface cation vacancy (SCV) structure \cite{Bliem14}. This revelation highlights the issues that might arise from generalizing results from surface-sensitive techniques, such as x-ray absorption spectroscopy (XAS) in total electron yield (TEY) mode and x-ray photoelectron spectroscopy (XPS), to explain the behavior of the bulk material \cite{Parkinson16}.

In particular, drawing conclusions about the cation distribution of magnetite requires caution, because the bulk material of the inverse spinel $\mathrm{Fe_3O_4}$ should contain divalent $\mathrm{Fe^{2+}_{oct}}$, as well as trivalent ions in both octahedral and tetrahedral coordination, $\mathrm{Fe^{3+}_{tet}}$ and $\mathrm{Fe^{3+}_{oct}}$. 
However, the DFT+U calculations of the SCV structure predict the first four atomic layers to only contain $\mathrm{Fe^{3+}}$ ions and to have a formal stoichometry of $\mathrm{Fe_{11}O_{16}}$, in agreement with earlier reports on an excess of $\mathrm{Fe^{3+}}$ at the (001) surface \cite{Chambers99}. A subsequent study of how the magnetic properties of $\mathrm{Fe_3O_4}$ are affected at the surface used x-ray magnetic circular dichroism (XMCD) with varying probing depth and found a reduced magnetic moment at the surface of a natural magnetite crystal \cite{Martin-Garcia15}.

In this letter we report an investigation of the magnetic surface properties of ultrathin $\mathrm{Fe_3O_4}(001)$ films, in contrast to the bulk, by recording magnetooptical depth profiles of the three cation species in $\mathrm{Fe_3O_4}$. Using photons with resonant energies, we employed XRMR to determine the magnetooptical depth profiles at the energies characteristic for $\mathrm{Fe^{2+}_{oct}}$, $\mathrm{Fe^{3+}_{tet}}$ and $\mathrm{Fe^{3+}_{oct}}$ in magnetite's $L_{2,3}$ XMCD spectrum individually for three ultrathin $\mathrm{Fe_3O_4/MgO(001)}$ films of varying thicknesses. 
We find a $\approx 3.9\,\mathrm{\r{A}}$ layer of enhanced magnetooptical absorption at the surface at the resonant energies of both $\mathrm{Fe^{3+}}$ species but not for $\mathrm{Fe^{2+}}$, suggesting an $\mathrm{Fe^{3+}}$-rich surface.

\textit{Experimental and theoretical details.}$-$We prepared the $\mathrm{Fe_3O_4/MgO(001)}$ samples in a multichamber ultra-high-vacuum system using reactive molecular beam epitaxy (RMBE). Their chemical composition and \mbox{$(\sqrt{2} \times \sqrt{2})\mathrm{R}45^\circ$} superstructure was confirmed by \textit{in-situ} XPS and low-energy electron diffraction (LEED), respectively. For details on the deposition and characterization methods please see Refs. \cite{Kuepper16} or \cite{Kuschel16}.

For the XAS and XMCD study, the samples were transfered from our lab under ambient conditions to the Superconducting Vector Magnet Endstation at beamline 4.0.2 of the Advanced Light Source (ALS). All samples were measured in a magnetic field of $4\,$T along the x-ray beam at room temperature. The incidence angle of the x-rays was \ang{30} from the [100] direction of $\mathrm{Fe_3O_4}$, and the degree of circular polarization was 90\%. The XAS and XMCD spectra were measured across the Fe $L_{2,3}$ absorption edges ($690 - 750\,\si{eV}$). All XAS spectra were measured in the TEY mode, which has a probing depth in magnetite of about $3\,$nm.

The XAS and XMCD data were analyzed by applying the sum rules \cite{Piamonteze09,Chen95,Teramura96} and charge-transfer multiplet calculations using the Thole code \cite{VanDerLaan97} with assistance of CTM4XAS \cite{deGroot05,Stavitski10}. For the sum rules, we took into account a correction factor of 1.142 derived by Teramura \textit{et al.} for $\mathrm{Fe^{2+}}$  \cite{Teramura96} and assumed $14\,\mathrm{\frac{holes}{f.u.}}$. For the multiplet calculations, we assumed the three-cation model, using crystal field and charge-transfer parameters as described in Ref. \cite{Kuepper16}. The parameters and more details regarding the multiplet calculations can also be found in the Supplemental Materials \cite{Supple} (Chapter A, including Refs. \cite{Regan01,Chantler95,Kuepper16}).
The multiplet states resulting from these calculations were compared to the data by assuming a Gaussian instrumental broadening of $0.2\,$eV, and a Lorentzian lifetime broadening of $0.3\,$eV at $L_3$ and $0.6\,$eV at $L_2$. 

The samples were transfered to BESSY II under ambient conditions and x-ray reflectivity (XRR) and XRMR were performed in the XUV Diffractometer at beamline $\text{UE46\_PGM-1}$  \cite{UE46PGM1}. The samples were placed between two permanent magnets in a magnetic field of $~200\,$mT at room temperature, with a degree of circular polarization of 90\%.
First, we characterized the structural properties of the sample (thickness $d$, roughness $\sigma$) by XRR at off-resonant energies ($680\,\si{eV}$, $1000\,\si{eV}$). Second, XAS and XMCD were measured in order to select suitable energies for XRMR. Finally, $\theta-2\theta$ scans in the range $2\theta=\ang{0} - \ang{140}$ at resonant energies $E_i$ with extrema in the XMCD signal (maximum at $708.4\,\si{eV}$, minimum at $709.5\,\si{eV}$, maximum at  $710.2\,\si{eV}$) were performed with both right and left circularly polarized x-rays, to obtain the XRMR asymmetry ratios $\Delta I(E_i,q_z)$. These curves were then fitted with the Zak matrix formalism using the software ReMagX \cite{Macke14} to determine the depth profiles of the magnetooptical absorption $\Delta \beta(z)$ and dispersion $\Delta \delta(z)$ along the film height $z$.
A detailed review of the XRMR method and the software is given in Ref. \cite{Macke14}, and a conclusive recipe for fitting XRMR data can be found in Refs. \cite{Kuschel15,Klewe16}.

\begin{table}[b]
\caption{\label{tab_thicc} Thicknesses $d$, surface roughnesses $\sigma^\mathrm{surf}$ and interface roughnesses $\sigma^\mathrm{int}$ of the investigated samples as obtained by fitting off-resonant XRR curves, as well as their spin moments $\mu_\mathrm{spin}$ and orbital moments $\mu_\mathrm{orb}$ from the sum-rule analysis.}
\begin{ruledtabular}
\begin{tabular}{c|ccc}
sample & $13\,$nm
& $25\,$nm & $50\,$nm
\\ 
\hline
$d$ (nm) & $13.5 \pm 0.5\,$ & $25.2 \pm 0.3\,$ & $52.8 \pm 0.3\,$ \\
$\sigma^\mathrm{surf}$ (nm) & $0.22 \pm 0.05\,$ & $0.33 \pm 0.05\,$ & $0.34 \pm 0.05\,$\\
$\sigma^\mathrm{int}$(nm) & $0.35 \pm 0.05\,$ & $0.35 \pm 0.05\,$ & $0.37 \pm 0.05\,$\\
\hline
$\mu_\mathrm{spin}~(\mathrm{\mu_B/f.u.})$ & $3.2 \pm 0.3$ & $3.5 \pm 0.3$ &  $3.7 \pm 0.3$\\
$\mu_\mathrm{orb}~(\mathrm{\mu_B/f.u.})$ & $0.07 \pm\,0.02$ & $0.09 \pm\,0.02$ &  $0.11 \pm 0.02$\\
$\mu ~(\mathrm{\mu_B/f.u.})$ & $3.27 \pm\,0.3$ & $3.59 \pm\,0.3$ &  $3.81 \pm 0.3$\\
\end{tabular} 
\end{ruledtabular}
\end{table}
\begin{figure}[t]
\centering
\includegraphics[scale=0.35]{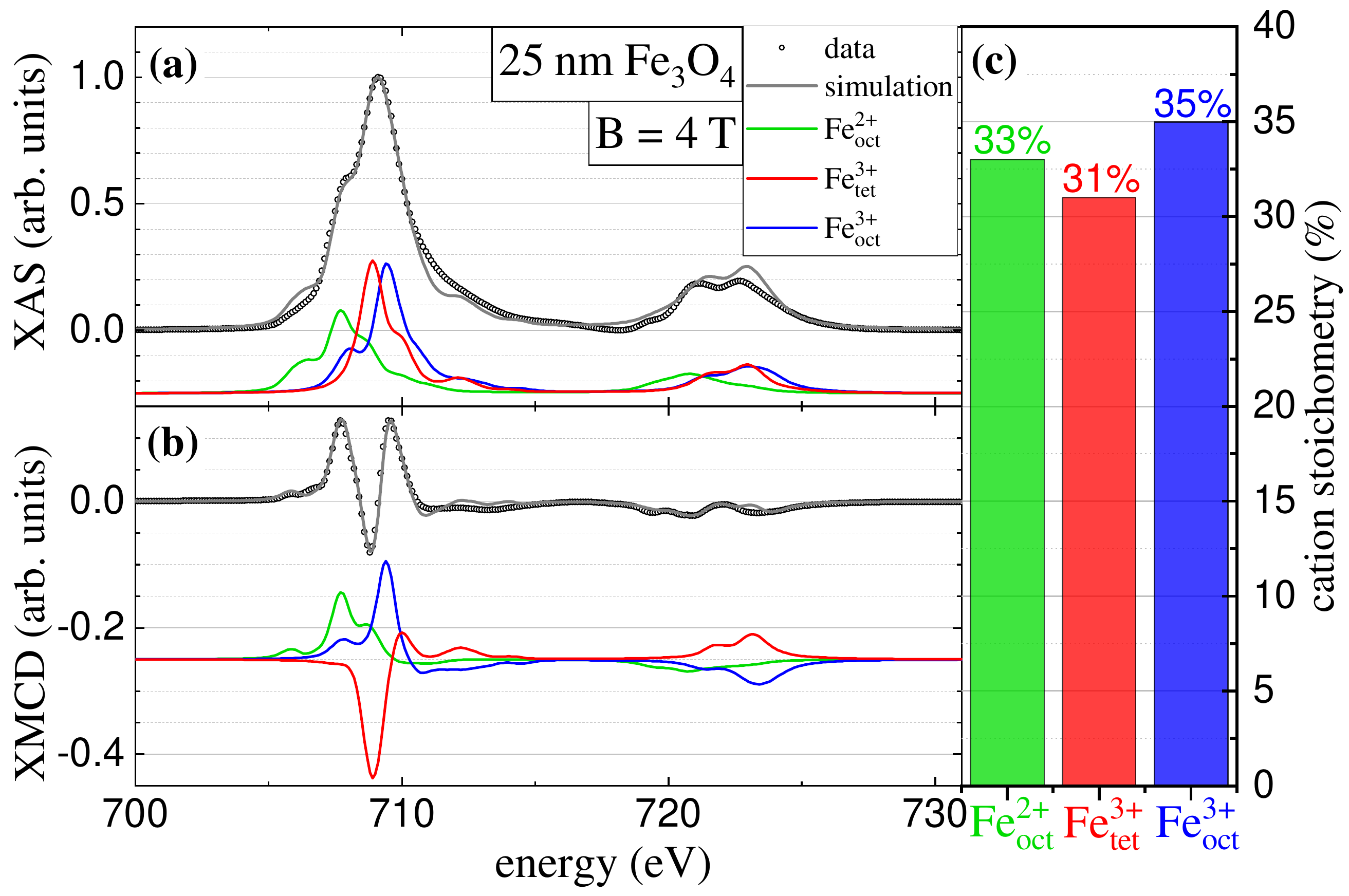}
\caption{(a) XAS and (b) XMCD spectrum at the Fe $L_{2,3}$ edge for the $25\,$nm $\mathrm{Fe_3O_4}$ film, taken at $4\,$T external magnetic field, at room temperature and in TEY mode. A step function was subtracted from the XAS spectrum. Black dots represent data; green, red and blue spectra are multiplet calculations for the three cation species of $\mathrm{Fe_3O_4}$, the grey line is their sum. The cation spectra are offset for better visibility. (c) Cation stoichometry used to obtain the fit in (a) and (b).}
\label{fig_XMCD}
\end{figure}
\begin{table}[b]
\caption{\label{tab_resonances} Contributions of the three cation species to the extrema in the XMCD spectrum in Fig. \ref{fig_XMCD}(b), as obtained by the multiplet analysis using Eq. (\ref{eq_contributions}).}
\begin{ruledtabular}
\begin{tabular}{cccc}
Energy & $\mathrm{Fe^{2+} _{oct}}$ & $\mathrm{Fe^{3+} _{tet}}$ & $\mathrm{Fe^{3+} _{oct}}$ \\ 
\hline
$708.4\,$eV & $73 \pm 5\%$ & $-8 \pm 3\%$ & $19 \pm 5\%$  \\ 
$709.5\,$eV & $18 \pm 3\%$ & $-64 \pm 3\%$ & $18\pm 3\%$ \\ 
$710.2\,$eV & $4\pm 3\%$ & $-16 \pm 8 \%$ & $80 \pm 10 \%$ \\ 
\end{tabular} 
\end{ruledtabular}
\end{table}
\textit{Results.}$-$Structural parameters obtained from the off-resonant XRR measurements are displayed in Tab. \ref{tab_thicc}.
Figure \ref{fig_XMCD} shows the XAS and XMCD spectra of the $25\,$nm $\mathrm{Fe_3O_4}$ film, recorded at ALS. Corresponding data measured at BESSY II, under the same conditions in which the XRMR was performed, can be found in the Supplemental Material \cite{Supple} (Chapter A).
The spin and orbital moments $\mu _\mathrm{spin}$ and $\mu_\mathrm{orb}$ obtained from the sum rules are given in Tab. \ref{tab_thicc}. Their sum is slightly reduced compared to the bulk value of magnetite of $\mu = \mu_\mathrm{spin} + \mu_\mathrm{orb} = 4.07\,\mu_\mathrm{B}$ \cite{Weiss29}. 
They also show a tendency toward a slightly higher moment for the thicker films. This behavior of magnetite films has also been observed previously \cite{Kuepper16}, and can be explained by a higher density of anti-phase boundaries (APBs) for thinner films due to the antiferromagnetic coupling of APBs reducing the average magnetic moment of the film \cite{Celotto03,Ramos06,Moussy04}.

Additionally, multiplet simulations of the three-cation model are fitted to the XMCD data (cf Fig. \ref{fig_XMCD}). By weighting the individual spectra with respect to the cation stoichometry given in Fig. \ref{fig_XMCD}(c), the XAS and XMCD data can be described well by our model (cf. grey line). Thus, the cation distribution on different sites almost follows the ideal stoichiometry of 1:1:1.
 
One feature of this kind of modelling is the fact that each of the three extrema observed in the XMCD spectrum can mainly be attributed to one cation spectrum. Table \ref{tab_resonances} shows the contributions $r^\mathrm{cation}(E_i)$ of each cation spectrum at the resonant energies $E_i$ in the XMCD spectrum, according to
\begin{equation}
r^\mathrm{cation}(E_i) = \frac{I^\mathrm{cation}(E_i)}{|I^\mathrm{Fe^{2+}_{oct}}(E_i)|+|I^\mathrm{Fe^{3+}_{tet}}(E_i)|+|I^\mathrm{Fe^{3+}_{oct}}(E_i)|}~~~,
\label{eq_contributions}
\end{equation}
with $I^\mathrm{cation}(E_i)$ being the XMCD signal of the corresponding cation spectrum in Fig. \ref{fig_XMCD}(b) at energies $E_i = 708.4\,\mathrm{eV}, 709.5\,\mathrm{eV}, 710.2\,\mathrm{eV}$.
While there still is a considerable mixing, at least $64\%$ of each extremum can be attributed to its dominant cation. 

The distinction between the $\mathrm{Fe^{2+}_{oct}}$ and $\mathrm{Fe^{3+}_{oct}}$ ions is a noteworthy issue for several reasons. Above the Verwey transition temperature, these ions should be randomly distributed on octahedral sites as shown by neutron diffraction \cite{Garcia00}. Thus, the model could potentially be simplified by describing them as $\mathrm{Fe^{2.5+}_{oct}}$, effectively reducing the number of cation species. However, only multiplet calculations based on a three-cation model describe room-temperature spectra from XPS, XAS and XMCD at the Fe $L$ absorption edge with sufficient agreement \cite{deGroot90,Kuiper97,Miedema13,Kuepper16,Phase19}. 
Note that this extremum-cation assignment qualitatively also holds true in LSDA+U calculations, but they predict a stronger overlap of the three cation spectra \cite{Antonov03}. A possible explanation for this behavior may well be the different electronic structure at the surface of magnetite. In that case, we would expect different spectra from the surface and the bulk. In fact, a recent study using hard x-ray photoelectron spectroscopy (HAXPES) reported on a bulk-exclusive state not observable in surface-sensitive soft XPS \cite{Taguchi15}.

Accordingly, the strategy to obtain cationic depth profiles is to pick the three corresponding XMCD resonant energies and perform XRMR measurements at these resonances.
\begin{figure}[t]
\centering
\includegraphics[scale=0.35]{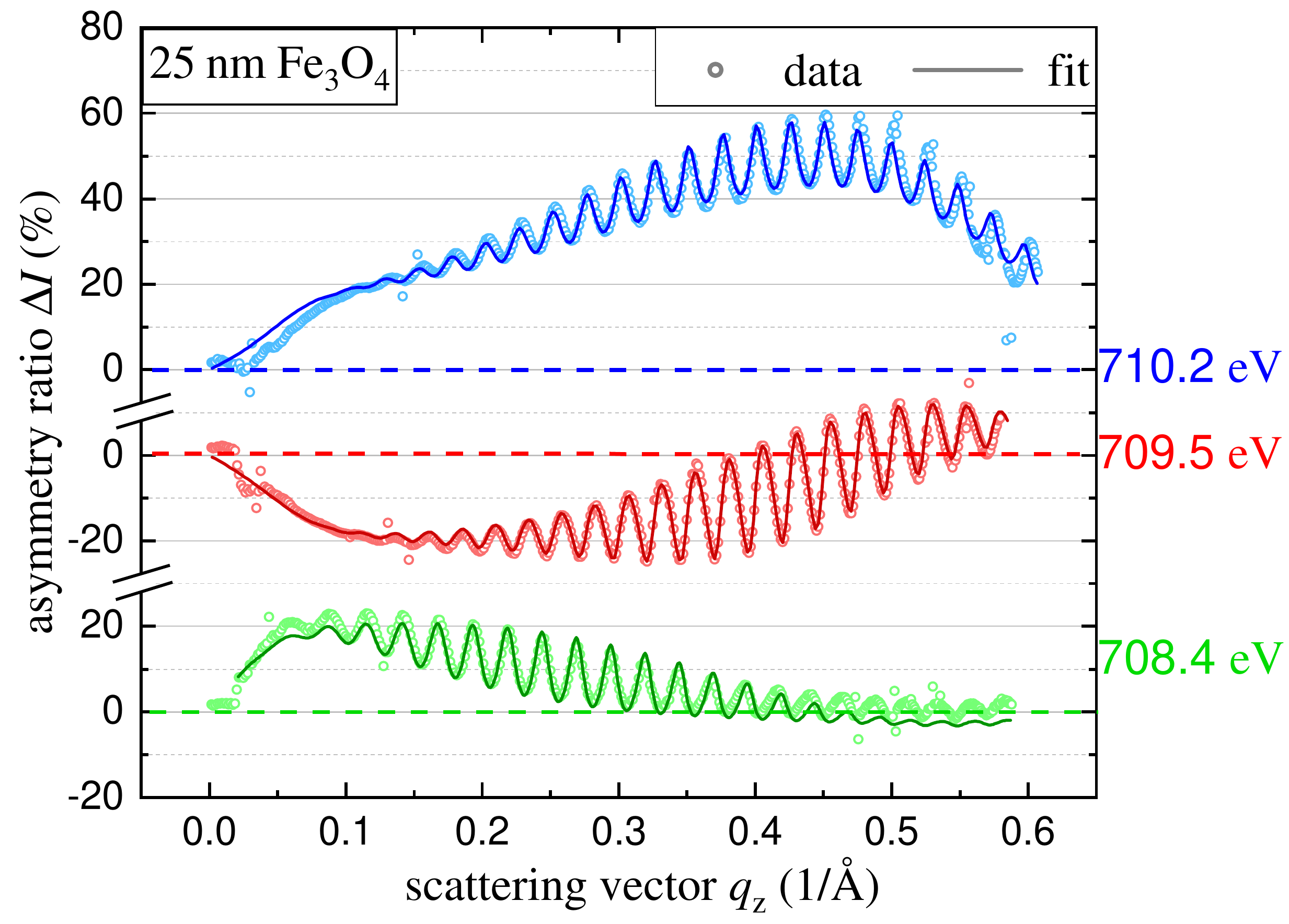}
\caption{XRMR data (open circles) and corresponding fits (solid lines) from the $25\,$nm $\mathrm{Fe_3O_4}$ film, recorded at the three resonant energies of the XMCD $L_3$ edge, using the modeled magnetooptical depth profiles of Fig. \ref{fig_Profiles}(a). Data were recorded with a magnetic field of $200\,$\si{mT} along the $\mathrm{Fe_3O_4}$(001) direction at room temperature.}
\label{fig_XRMR}
\end{figure}
\begin{figure}[t]
\centering
\includegraphics[scale=0.35]{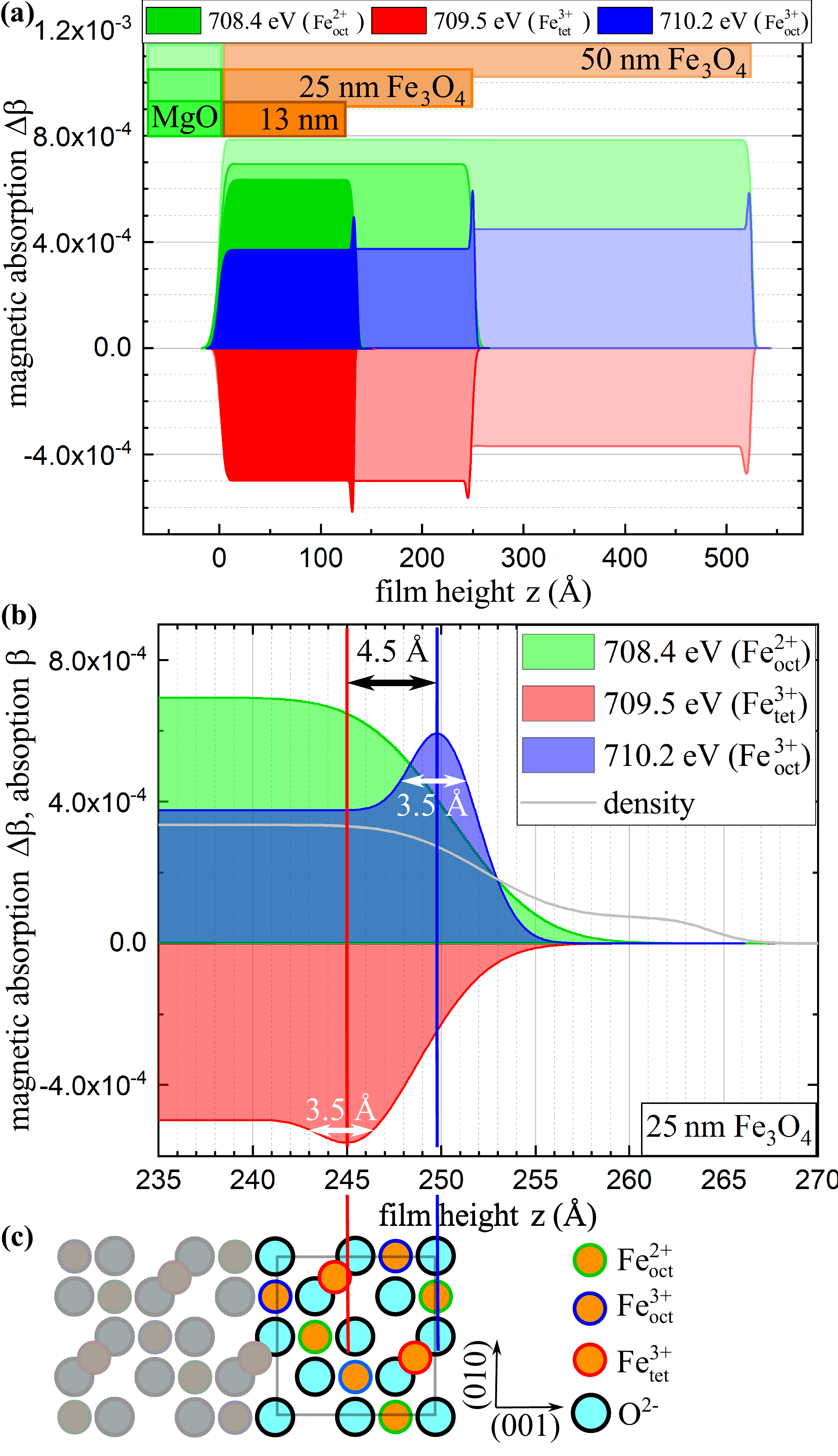}
\caption{(a) $\Delta \beta (z)$ depth profiles for all three samples at the three resonant energies, extracted from the XRMR fits. (b) Close-up of the surface magnetooptical depth profile of the $25\,$nm $\mathrm{Fe_3O_4}$ film, together with the optical density obtained from off-resonant XRR fits (grey line). (c) (Bulk-terminated) model of the magnetite unit cell, in scale with Fig. 3(b). Comparison with the model in (c) illustrates the sizes of the enhanced regions being roughly half a unit cell of magnetite (four cation layers).}
\label{fig_Profiles}
\end{figure}
Figure \ref{fig_XRMR} shows the asymmetry ratios and their fits at the three XMCD resonant energies for the $25\,$nm $\mathrm{Fe_3O_4}$ film. Corresponding figures for the other two samples can be found in the Supplemental Material \cite{Supple} (Chapter B). 
The magnetooptical depth profiles $\Delta \beta(z)$ that produce the fits are shown in Fig. \ref{fig_Profiles}(a).   
The most striking feature of all three samples is the behavior at the surface: at the $\mathrm{Fe^{2+} _{oct}}$ resonance energy ($708.4\,$eV, green), the magnetooptical depth profiles in fact appear to be just homogeneous for all samples. However, at both the $\mathrm{Fe^{3+}_{tet}}$ and the $\mathrm{Fe^{3+}_{oct}}$ resonance energies, there are noticeable changes to the $\Delta \beta$ depth profiles. In order to fit their asymmetry ratios, we must include a thin surface layer of enhanced magnetooptical absorption. The two thinner $\mathrm{Fe_3O_4}$ films - $13\,$nm and $25\,$nm - are quantitatively very similar for the $\mathrm{Fe^{3+} _{oct}}$ and $\mathrm{Fe^{3+} _{tet}}$ magnetooptical depth profiles, with only minor differences in the enhanced amplitude at the surface, and the $\Delta \beta$ in the bulk matches between the samples. In contrast, the $50\,$nm $\mathrm{Fe_3O_4}$ film shows slightly higher $\Delta \beta$ at the $\mathrm{Fe^{3+} _{oct}}$ resonance, and smaller $\Delta \beta$ at the $\mathrm{Fe^{3+} _{tet}}$ resonance. Also, the magnetooptical absorption at the $\mathrm{Fe^{2+} _{oct}}$ resonance becomes larger with increasing film thickness, in agreement with our results from the sum-rule analysis (see Tab. \ref{tab_thicc}).
The obvious choice of magnetooptical depth profiles which are simply homogeneous through the entire film did not provide satisfactory fits to the data. This is discussed in more detail in the Supplemental Material \cite{Supple} (Chapter C).

In order to highlight this phenomenon, Fig. \ref{fig_Profiles}(b) shows the surface region of the $25\,$nm $\mathrm{Fe_3O_4}$ film, together with the density depth profile obtained from off-resonant XRR (grey line).
The edge of the magnetooptical depth profile of the $\mathrm{Fe^{2+} _{oct}}$ resonance roughly matches the location of the magnetically enhanced $\mathrm{Fe^{3+}_{tet}}$ layer. The thickness of the magnetically enhanced layers is about $3.5\,\mathrm{\r{A}}$ for both $\mathrm{Fe^{3+}}$ species. This corresponds to slightly less than half a bulk unit cell of magnetite ($a/2=4.2\,\mathrm{\r{A}}$), as illustrated by Fig. \ref{fig_Profiles}(c).
Furthermore, the magnetically enhanced layers are not colocated at the same depth: the magnetically enhanced $\mathrm{Fe^{3+} _{tet}}$ layer is shifted about $4.5\,$\r{A} deeper into the film than the magnetically enhanced $\mathrm{Fe^{3+} _{oct}}$ layer. Table \ref{tab_profiles} summarizes the individual thicknesses of the magnetically enhanced layers $d^\mathrm{enh}$ and their offset from one another for the different samples. All three samples show comparable, but not quite identical results. While the model qualitatively holds up well among the samples, both the thicknesses $d^\mathrm{enh}$ and the depth offsets are similar in magnitude to the surface roughness, making it difficult to resolve the exact distances with any greater precision.
\begin{table}[t]
\caption{\label{tab_profiles} Thicknesses $d^\mathrm{enh}$ and vertical shifts of the enhanced magnetic layers at the $\mathrm{Fe^{3+}_{tet}}$ and $\mathrm{Fe^{3+}_{oct}}$ resonances.}
\begin{ruledtabular}
\begin{tabular}{cccc}
• & $d^\mathrm{enh} _{\mathrm{Fe^{3+}_{oct}}}$ & $d^\mathrm{enh} _{\mathrm{Fe^{3+}_{tet}}}$ & vertical shift \\
\hline 
$13\,$nm & $4.7\,$\r{A} & $3.2\,$\r{A}& $2\,$\r{A} \\ 
$25\,$nm & $3.5\,$\r{A} & $3.4\,$\r{A} & $4.5\,$\r{A}\\ 
$50\,$nm & $3.8\,$\r{A} & $6.4\,$\r{A}& $2.5\,$\r{A}\\ 
\end{tabular} 
\end{ruledtabular}
\end{table}

\textit{Discussion.}$-$Within the three-cation picture, we can discuss the magnetooptical depth profiles obtained for the resonant energies.
The magnetooptical depth profiles are not identical with the depth distribution of the cations:
As quantified in Tab. \ref{tab_resonances}, however, the signal on each resonance is a mixture of contributions from all three cations. For the magnetooptical depth profile at $710.2\,$eV approximately $80\%$ of the signal originates from the $\mathrm{Fe^{3+}_{oct}}$ and can be regarded as an almost pure effect from that species. And since the position of the layer of enhanced magnetization at $709.5\,$eV does not match the position of the $710.2\,$eV layer, we can conclude it to be a distinct physical feature, stemming from the $\mathrm{Fe^{3+}_{tet}}$ species. 

One ansatz is to take into account rearranged cation distributions due to the $\mathrm{Fe_3O_4}(001)$ surface as proposed by the SCV model \cite{Bliem14, Parkinson16}. The SCV model predicts that, in order to achieve polarity compensation, the first unit cell contains only $\mathrm{Fe^{3+}}$ species, with the first $\mathrm{Fe^{3+}_{tet}}$ layer lying about 1 \r{A} deeper than the $\mathrm{Fe^{3+}_{oct}}$. This model matches surprisingly well some aspects of our findings. The first $\mathrm{Fe_{oct}}$-O layer remains stoichometric, but the $\mathrm{Fe^{2+}_{oct}}$ changes valency to $\mathrm{Fe^{3+}_{oct}}$, effectively doubling the $\mathrm{Fe^{3+}_{oct}}$ density. In the second layer, an additional $\mathrm{Fe^{3+}_{tet}}$ ion is added, increasing the $\mathrm{Fe^{3+}_{tet}}$ density by $50\%$. 
However, we do not observe the depletion of $\mathrm{Fe^{2+}_{oct}}$ cations in the first $8.4\,\mathrm{\r{A}}$.
This agreement is surprising because $\mathrm{Fe_3O_4}$ surfaces tend to hydroxilate on ambient conditions and do not show the $(\sqrt{2} \times \sqrt{2})\mathrm{R}45^\circ$ LEED pattern, but instead a $(1 \times 1)$ pattern \cite{Parkinson16}. This, however, may be attributed to disorder at the surface with loss of long-range order while the local order of vacancies and interstitials is kept.
Our results now suggest that at least the $\mathrm{Fe^{3+}}$-enrichment of the surface remains intact under ambient conditions. A more detailed comparison of the SCV model to our findings can be found in the Supplemental Material \cite{Supple} (Chapter D).

Taking the model of occupation of octahedral sites by $\mathrm{Fe^{2.5+}_{oct}}$ cations as an alternative, the agreement of the XAS/XMCD spectra with the multiplet calculations might be merely valid at the surface, while both octahedrally coordinated Fe species are identical in the bulk. In that case, the discrepancy between surface and bulk of the magnetooptical depth profiles would represent the transition from the surface electronic structure to the bulk structure. Using bulk-sensitive HAXPES, Taguchi \textit{et al.} report on a bulk feature at $708.5\,$eV, which is invisible for surface-sensitive soft XPS \cite{Taguchi15}. 
In this picture, we could interpret the magnetooptical depth profiles and the XMCD spectra as follows: the top $4-6\,$\r{A}, consisting of $\mathrm{Fe^{3+}_{oct}}$ and $\mathrm{Fe^{3+}_{tet}}$ ions, give rise to the XMCD extrema at $709.5\,$eV and $710.2\,$eV, as reflected in the enhanced $\Delta \beta$ layers at those energies. Deeper into the $\mathrm{Fe_3O_4}$ film, the $\mathrm{Fe^{3+}_{oct}}$ species vanishes in favor of the $\mathrm{Fe_{oct} ^{2.5+}}$ species, and consequently, the magnetic dichroism at $710.2\,$eV is reduced. Instead, the bulk-feature at $708.4\,$eV becomes prominent, related to the $\mathrm{Fe_{oct} ^{2.5+}}$ species. In the XMCD signal, it is visible as the supposed $\mathrm{Fe^{2+}_{oct}}$ peak, but strongly suppressed due to the surface sensitivity of the TEY detection. In this interpretation, our results agree with the more pronounced signal at $708.5\,$eV far from the surface reported in Ref. \cite{Taguchi15}, and can explain the observation of the two distinct $\mathrm{Fe_{oct}}$ species at the surface, which should not be distinguishable at room temperature.

\textit{Summary.}$-$In conclusion, we fabricated three $\mathrm{Fe_3O_4 /MgO(001)}$ ultrathin films of different thicknesses by RMBE. We recorded XAS/XMCD at ALS beamline 4.0.2 as well as XAS/XMCD and XRMR measurements at BESSY II beamline $\mathrm{UE46\_PGM}$-1, in order to obtain magnetooptical depth profiles. By fitting multiplet calculations to the XMCD data, we disentangle the cation contributions at the three resonant energies of the XMCD spectrum, and use XRMR at those energies in order to resolve the magnetooptical depth profiles of the three iron species in $\mathrm{Fe_3O_4}$. We find that both $\mathrm{Fe^{3+}}$ species show an enhanced signal in the surface-near region in a $\approx 3.9 \pm 1$ \r{A} thick layer, with the $\mathrm{Fe^{3+}_{tet}}$ layer located about $3 \pm 1.5$ \r{A} underneath the $\mathrm{Fe^{3+}_{oct}}$ layer. We attribute this to the first unit cell from the surface containing an excess of $\mathrm{Fe^{3+}}$ cations. 

\textit{Acknowledgements.}$-$Financial support from the Bundesministerium f\"ur Bildung und Forschung (FKZ 05K16MP1) is gratefully acknowledged.  We are also grateful for the kind support from the Deutsche Forschungsgemeinschaft (DFG under  No. KU2321/6-1, and No. WO533/20-1).
This research used resources of the Advanced Light Source, a DOE Office of Science User Facility under contract no. DE-AC02-05CH11231, by recording XMCD at at beamline 4.0.2. (ALS-10261).
We thank HZB for the allocation of synchrotron beamtime at beamline $\mathrm{UE46\_PGM}$-1 $\text{(181/06266ST/R)}$ where we recorded the XRMR and additional XMCD measurements.

\bibliography{CationProfilesBib}
\end{document}


\title{Cation- and lattice-site-selective magnetic depth profiles of ultrathin $\mathbf{Fe_3O_4(001)}$ films}

\begin{center}
\large{Supplementary Information}\\
\begin{LARGE}
\textbf{Cation- and lattice-site-selective magnetic depth profiles of ultrathin $\mathbf{Fe_3O_4(001)}$ films}\\~\\
\end{LARGE}
\large{Tobias Pohlmann, Timo Kuschel, Jari Rodewald, Jannis Thien, Kevin Ruwisch, Florian Bertram, Eugen Weschke, Padraic Shafer, Joachim Wollschl\"{a}ger and Karsten K\"{u}pper}
\end{center}

\section{XMCD data from BESSY-II}
\begin{figure}[h]
\centering
\includegraphics[scale=0.35]{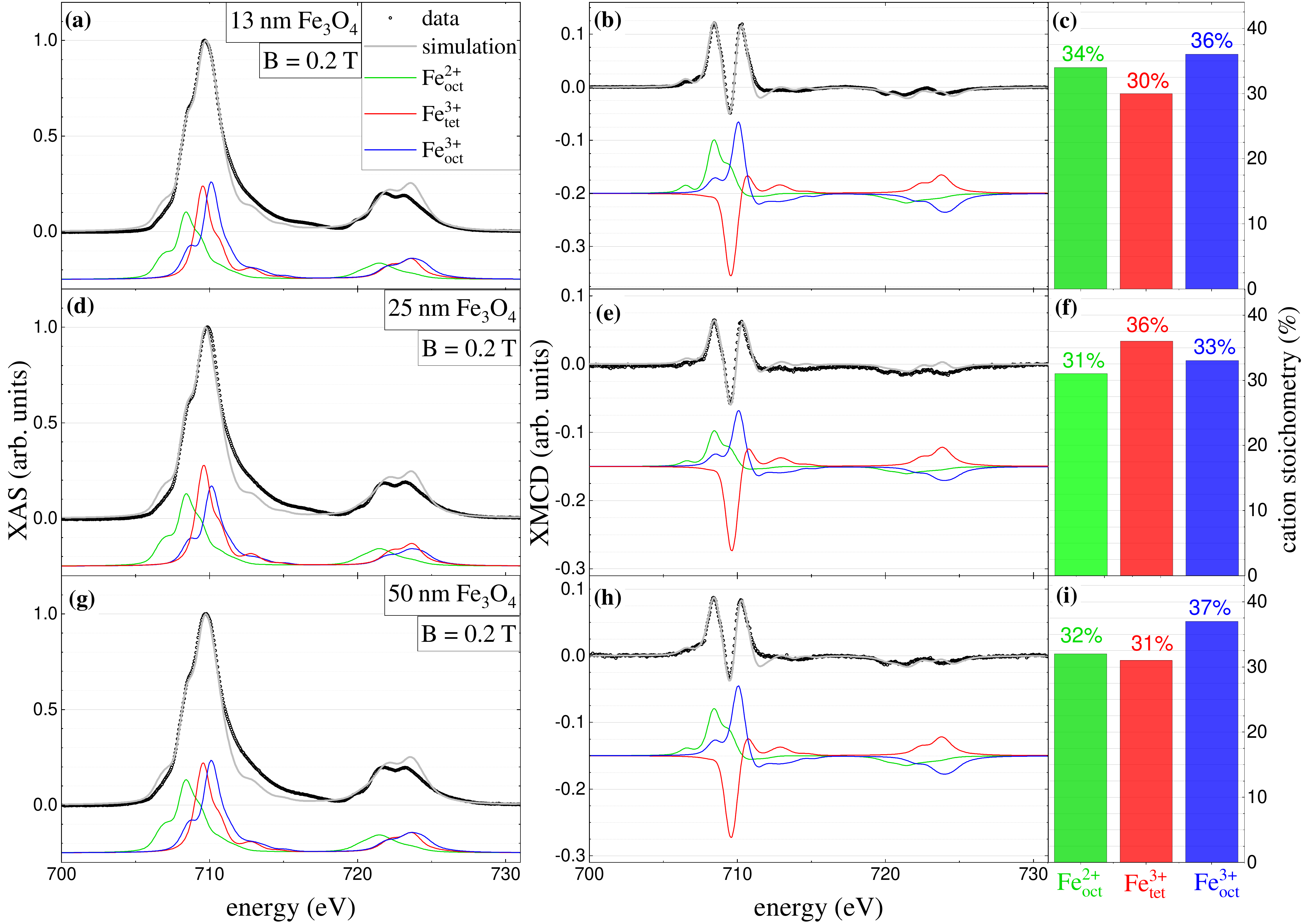}
\caption{XAS and XMCD spectra together with multiplet simulations and the extracted cation stoichometry for (a)-(c) the $13\,$nm $\mathrm{Fe_3O_4}$ film, (d)-(f) the $25\,$nm $\mathrm{Fe_3O_4}$ film and (g)-(i) the $50\,$nm $\mathrm{Fe_3O_4}$ film, recorded immediately before the XRMR measurements at BESSY, in a magnetic field of about $B \approx 200\,$mT at the sample location. The multiplet calculations were done using the parameters in Tab. \ref{tab:mp}.}
\label{XMCD}
\end{figure}
The XAS/XMCD data displayed in Fig. 1 of the main text were recorded at ALS in a $4\,$T vector magnet, making sure to saturate the sample magnetization for a reliable sum-rule evaluation. Also, since this endstation is dedicated to and optimized for XMCD, the data have better signal-to-noise ratio than those taken at $\mathrm{UE46\_PGM}$-1 of BESSY II, which specializes in resonant diffraction. For the XRMR measurements at $\mathrm{UE46\_PGM}$-1, the samples were located in between two permanent magnets with a magnetic field of about $200\,$mT at the sample location. We recorded there XAS/XMCD by the TEY detection scheme as well, in order to identify the energies of the XMCD extrema. These XAS and XMCD measurements, which were recorded alongside the XRMR, together with the multiplet fits and the resulting cation stoichometry, are plotted in Fig. \ref{XMCD}(a)-(c) for the $13\,$nm $\mathrm{Fe_3O_4}$ film, in Fig. \ref{XMCD}(d)-(f) for the $25\,$nm $\mathrm{Fe_3O_4}$ film and in Fig. \ref{XMCD}(g)-(i) for the $50\,$nm $\mathrm{Fe_3O_4}$ film. We measured XAS spectra with both right and left circularly polarized x-rays, $\mathrm{XAS^{right}}$ and $\mathrm{XAS^{left}}$. The non-dichroic XAS spectra shown in this work are the sum of these two XAS spectra, and the XMCD spectra their difference:
\[
\mathrm{XAS} = \mathrm{XAS^{right}+XAS^{left}}
~~~~~~~~~
\mathrm{XMCD} = \mathrm{XAS^{right}-XAS^{left}}
\]
Both are normalized so that the XAS maximum equals 1.

TEY data often suffer from saturation effects, when the absorption cross-section $\mathrm{\mu_A}(E)$ directly at the $L_{2,3}$ edge becomes similar to the inverse electron escape depth $1/\lambda_\mathrm{e}$. These difficulties in our data are corrected by the method described in Ref. \cite{Regan01}: the edge jump of the data is scaled to literature values \cite{Chantler95} in order to obtain
the saturation-affected $\mu_\mathrm{A}'(E)$, and then
\begin{equation}
\mathrm{\mu_A}(E) = (\frac{1}{\mathrm{\mu_A'}(E)}-\lambda_\mathrm{e})^{-1}
\end{equation}
is used to obtain the saturation-corrected absorption cross-section $\mathrm{\mu_A}(E)$. We used $\lambda_\mathrm{e} = 30\,\mathrm{\r{A}}$, which was obtained from the damping of the Ni $L_{2,3}$ TEY signal in $\mathrm{Fe_3O_4/NiO/MgO}$ bilayer samples. After
that, a step function has been subtracted from the XAS spectra in order to compare it to the multiplet simulations. 

Multiplet fits were done using the same parameter set for both the ALS and the BESSY II data. We assumed the three-cation model, with crystal field energies of \mbox{$\mathrm{10Dq^{oct}}$} in octahedral and \mbox{$\mathrm{10Dq^{tet}}$} in tetrahedral coordination, and the splittings between the initial and final charge-transfer states \mbox{$\Delta_\mathrm{init}$} and \mbox{$\Delta _\mathrm{final}$}. The exchange splitting $g\cdot \mu_\mathrm{B}$ had to be set individually. They are the same ones as in Ref. \cite{Kuepper16}, but can also be found in Tab. \ref{tab:mp}. In order to compare the calculations with the data, a Gaussian instrument broadening of $0.2\,$eV, and a Lorentzian lifetime broadening of $0.3\,$eV for $L_3$ and $0.6\,$eV for $L_2$ were assumed. The exchange field $g\cdot \mu_\mathrm{B}$ of the simulations is smaller for the data taken at BESSY II compared to those from ALS. This indicates that our samples could not be fully saturated in a magnetic field of $200\,\mathrm{mT}$ used for the experiments at BESSY II. 

Within the accuracy of the method, the cation stoichometry is very much the same for all three samples. This means that the magnetically enhanced layer for the two $\mathrm{Fe^{3+}}$ species is not well observable by this evaluation. Instead we observe a cation stoichometry very close to the ideal 1:1:1 ratio expected for $\mathrm{Fe_3O_4}$.

\begin{center}
\begin{table}[h]
\centering
\caption{\label{tab:mp} Parameters used for the multiplet calculation. For comparison with the data, a Gaussian instrument broadening of $0.2\,$eV, and a Lorentzian lifetime broadening of $0.3\,$eV for $\mathrm{L_3}$ and $0.6\,$eV for $\mathrm{L_2}$ were assumed.}
\begin{tabular}{c|cc|cc|cc}
• & $\mathrm{10Dq^{oct}}$ & $\mathrm{10Dq^{tet}}$ & $\Delta_\mathrm{init}$ & $\Delta_\mathrm{final}$ & $g\cdot \mu_\mathrm{B}$ & $g\cdot \mu_\mathrm{B}$\\
• &• &• &• &• & (BESSY II) & (ALS)
\\ 
\hline 
$13\,$nm & $1.0\,$eV & $-0.6\,$eV & $6\,$eV & $5\,$eV & $0.028\,$eV & $0.04\,$eV \\ 
$25\,$nm & $1.0\,$eV & $-0.6\,$eV & $6\,$eV & $5\,$eV & $0.011\,$eV & $0.04\,$eV \\ 
$50\,$nm & $1.0\,$eV & $-0.6\,$eV & $6\,$eV & $5\,$eV & $0.016\,$eV & $0.04\,$eV \\
\end{tabular}
\end{table}
\end{center}
\newpage
\section{XRR/XRMR studies on $\mathbf{13\,}$nm and $\mathbf{50\,}$nm $\mathrm{Fe_3O_4}$ films}
\begin{figure}[h]
\centering
\subfloat{
\includegraphics[scale=0.38]{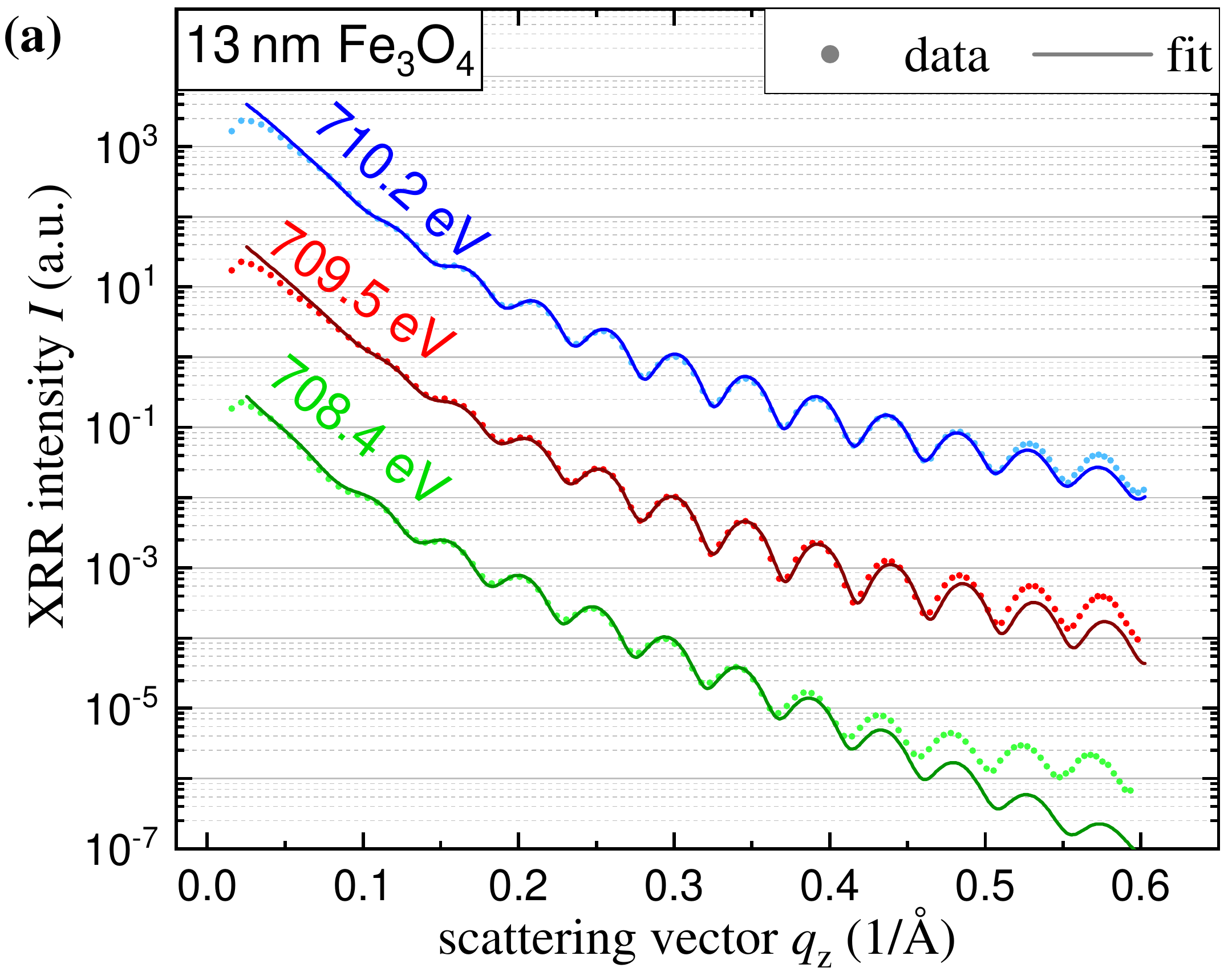}
}\hfill
\subfloat{\includegraphics[scale=0.38]{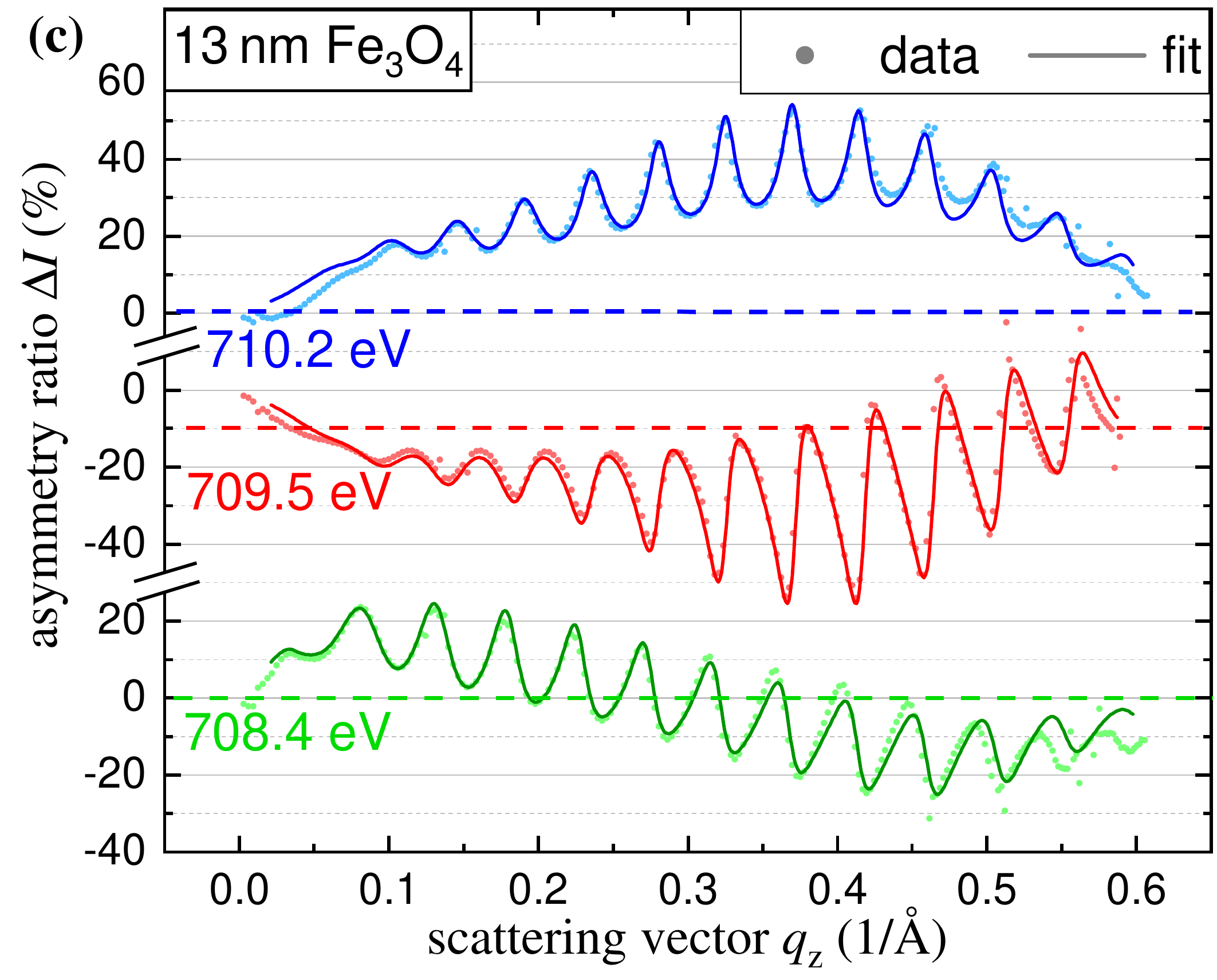}
}\\
\subfloat{\includegraphics[scale=0.38]{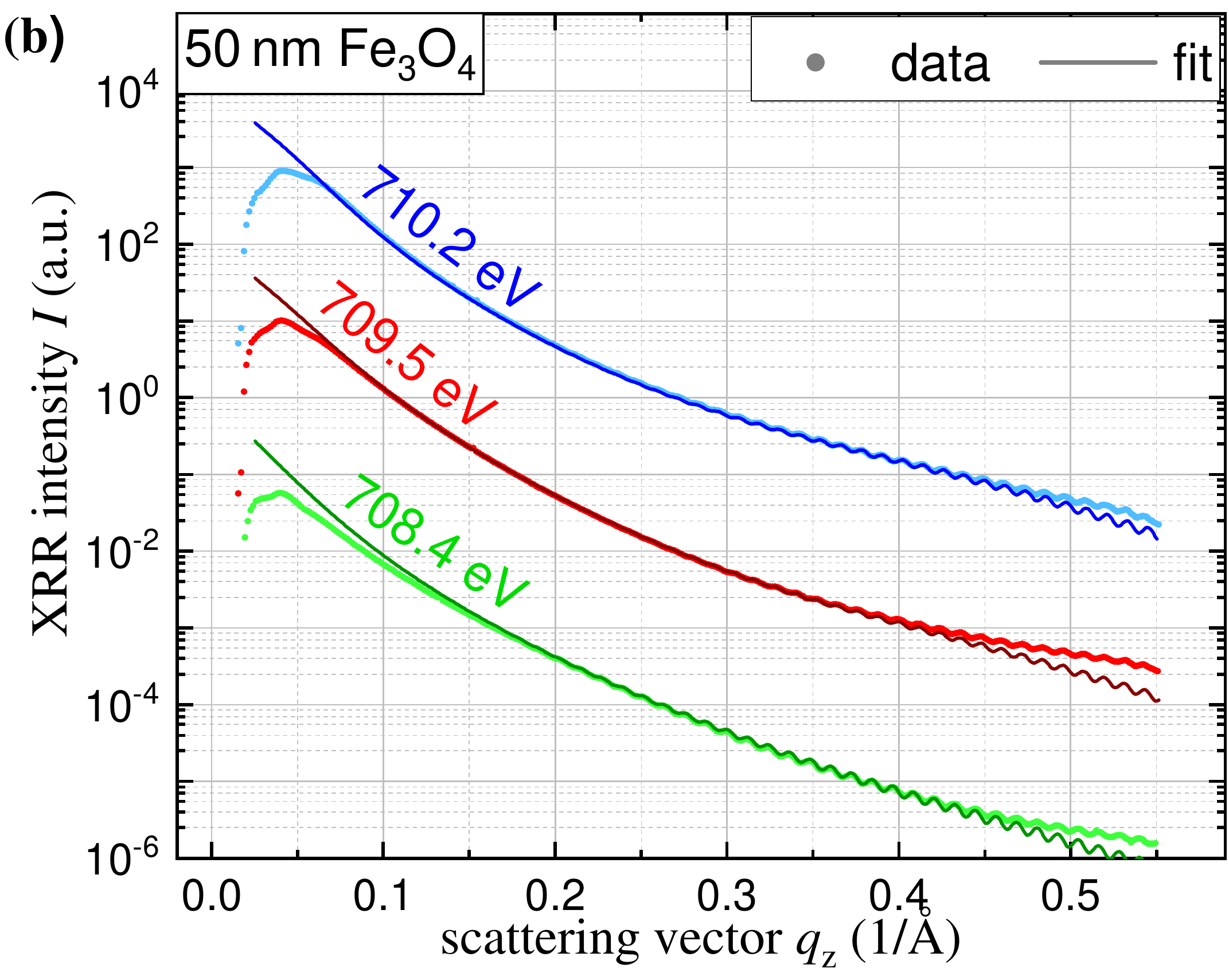}
}
\hfill
\subfloat{\includegraphics[scale=0.38]{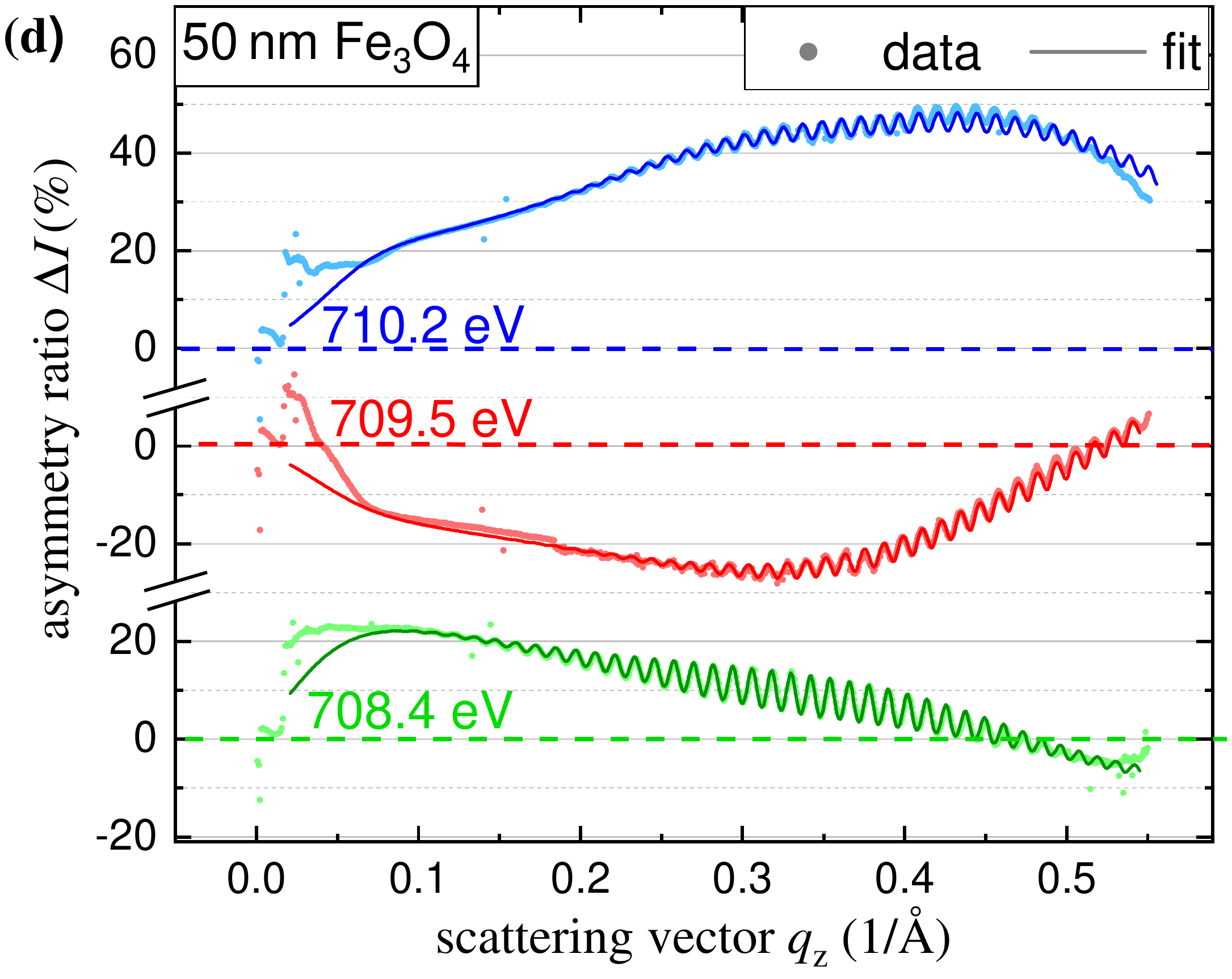}
}
\caption{XRR and XRMR data for the $13\,$nm and $50\,$nm $\mathrm{Fe_3O_4}$ films. The non-dichroic XRR curves with their fits are shown for the $13\,$nm $\mathrm{Fe_3O_4}$ film in (a) and for the $50\,$nm $\mathrm{Fe_3O_4}$ film in (b), and the XRMR asymmetry ratios together with fits are displayed for the $13\,$nm in c) and for the $50\,$nm $\mathrm{Fe_3O_4}$ film in d). The asymmetry fits are obtained by using the magnetooptical depth profiles in Fig. 3(a) in the main text.}
\label{1350}
\end{figure}

Figure \ref{1350} displays XRR and XRMR data for the $13\,$nm and $50\,$nm $\mathrm{Fe_3O_4}$ films, which were not included in the main text, where these data were only discussed exemplarily for the $25\,$nm $\mathrm{Fe_3O_4}$ film. Figures \ref{1350}(a) and (b) show the non-dichroic, resonant XRR data for the $13\,$nm and $50\,$nm $\mathrm{Fe_3O_4}$ films, respectively. Figures \ref{1350}(c) and (d) show the corresponding XRMR asymmetry ratios for those samples, together with the fits obtained from the magnetooptical depth profiles in Fig. 3(a) in the main text.
Experimental data are fitted well except for small deviations for high scattering vectors, when the reflectivity becomes weak and the signal counts become comparable to the dark current of the photodiode detector.
Overall, the model assuming top layers of enhanced magnetooptical absorptions works very well in fitting all XRMR data we recorded.
\newpage

\section{XRMR fits with homogeneous magnetization}

\begin{figure}[h]
\centering
\subfloat{
\includegraphics[scale=0.39]{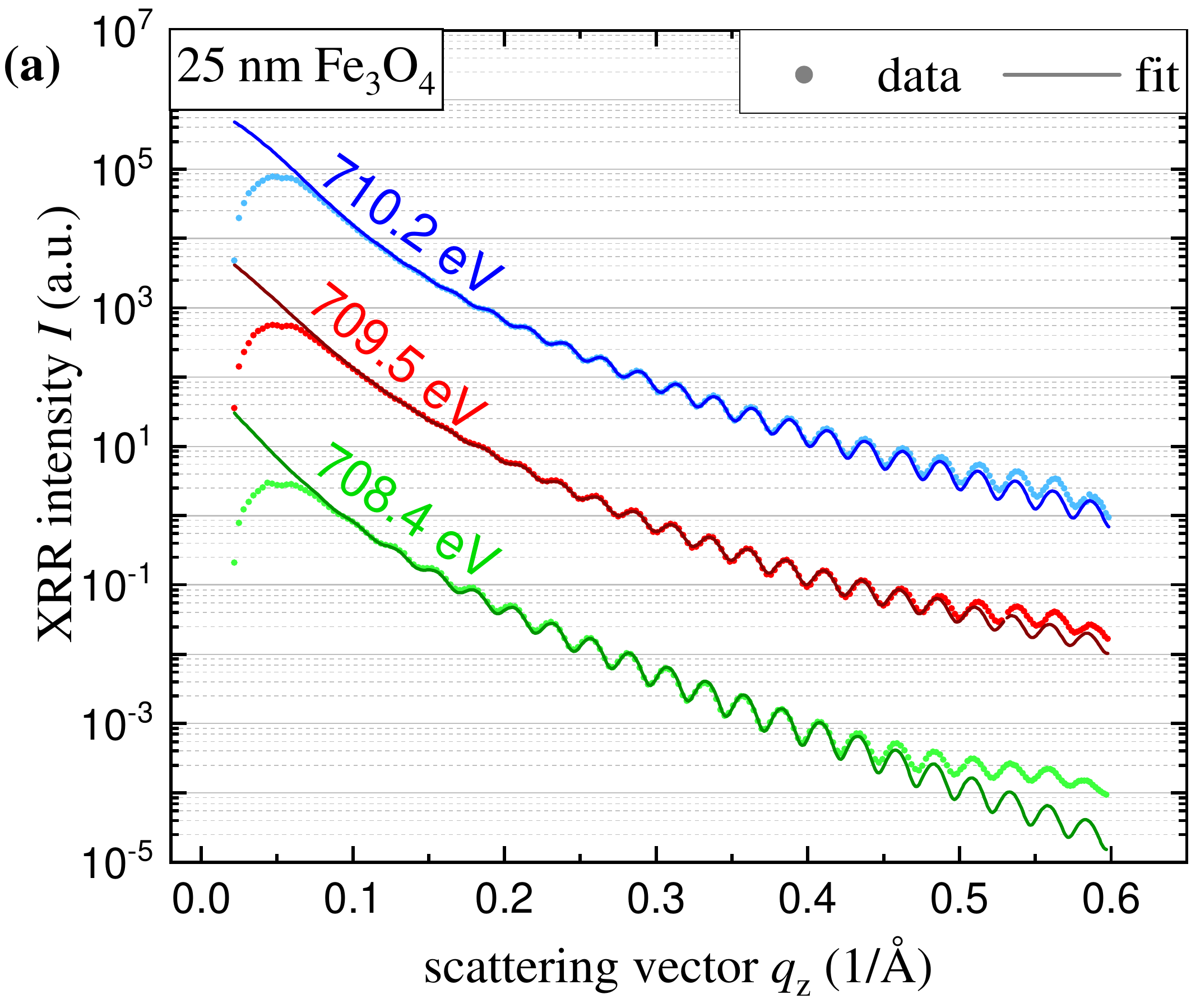}
}\hfill
\subfloat{
\includegraphics[scale=0.39]{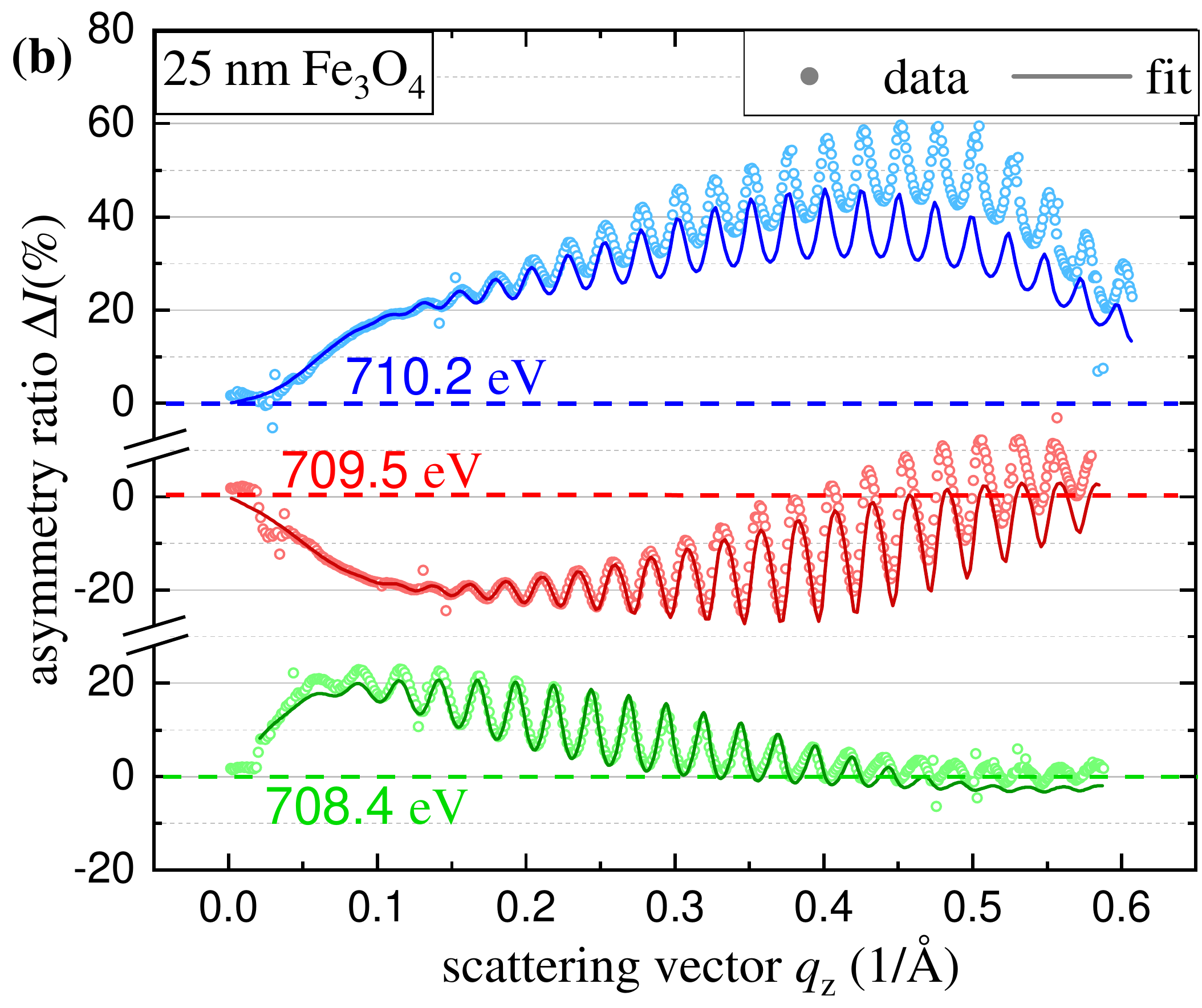}
}
\caption{(a) Non-dichroic XRR curves together with fits for the $25\,$nm $\mathrm{Fe_3O_4}$ film, corresponding to the XRMR asymmetry ratios shown in Fig. 2 in the main text. (b) Asymmetry ratios of the $25\,$nm $\mathrm{Fe_3O_4}$ film with the best fits achieved under the assumption of a single homogeneous magnetooptical profile. The curves at photon energies of $709.5\,$eV and $710.2\,$eV cannot be well accounted for already at scattering vectors $> 0.25 \mathrm{\,\r{A}^{-1}}$ without assuming a top layer of enhanced magnetooptical absorption (cf. Fig. 2 in the main text).}
\label{Homo}
\end{figure}

Figure \ref{Homo}(a) shows the resonant, non-dichroic XRR curves of the $25\,$nm $\mathrm{Fe_3O_4}$ film together with their fits, which are the base for the asymmetry ratio fits in Fig. 2 in the main text. Non-dichroic XRR curves are obtained by averaging the XRR curves recorded with right and left circularly polarized x-rays, $I = (I^\mathrm{right} + I^\mathrm{left})/2$.
The data are fitted well using a two-layer model, representing the $\mathrm{Fe_3O_4}$ film and an optically thin $0.5-1\,$nm top layer, which is likely due to adsorbants on the surface, as obtained from non-resonant XRR, and again only become inaccurate for high scattering vectors. It has to be noted that at resonant energy and therefore high absorption, the XRR curves are dominated by the optical absorption of the film, while all other fit parameters have only small impact on the XRR curve. Consequently, these fits themselves do not provide a very exact insight in the structural details of the samples. For those, we recorded off-resonant XRR.

In order to strengthen our claim of a top layer of enhanced magnetization, Fig. \ref{Homo}(b) shows again the asymmetry ratios already seen in Fig. 2 of the main text, but this time together with fits assuming homogeneous magnetization depth profiles, without the top layer of enhanced magnetization. These fits are the best we could obtain by allowing variation of the magneooptical constants $\Delta \beta, \Delta \delta$, individual roughnesses of the magnetooptical depth profiles at the substrate interface and the surface, as well as magnetic dead layers at both the subtrate interface and the surface. While the fit for the $708.4\,$eV curve does hold up very well, demonstrating a homogeneous distribution of $\mathrm{Fe^{2+}_{oct}}$ cations, the other two fits - especially at $710.2\,$eV - cannot describe the data satisfyingly. In order to account for these behaviours, we need to assume the layer of magnetooptical enhancement at the surface (see Fig. 2 in the main text). The fact that the XRMR curves of all three measured samples can be fitted well with this assumption makes this a very convincing model.  
\newpage
\section{Surface cation vacancy structure}
\begin{figure}[h]
\centering
\subfloat{
\includegraphics[scale=0.38]{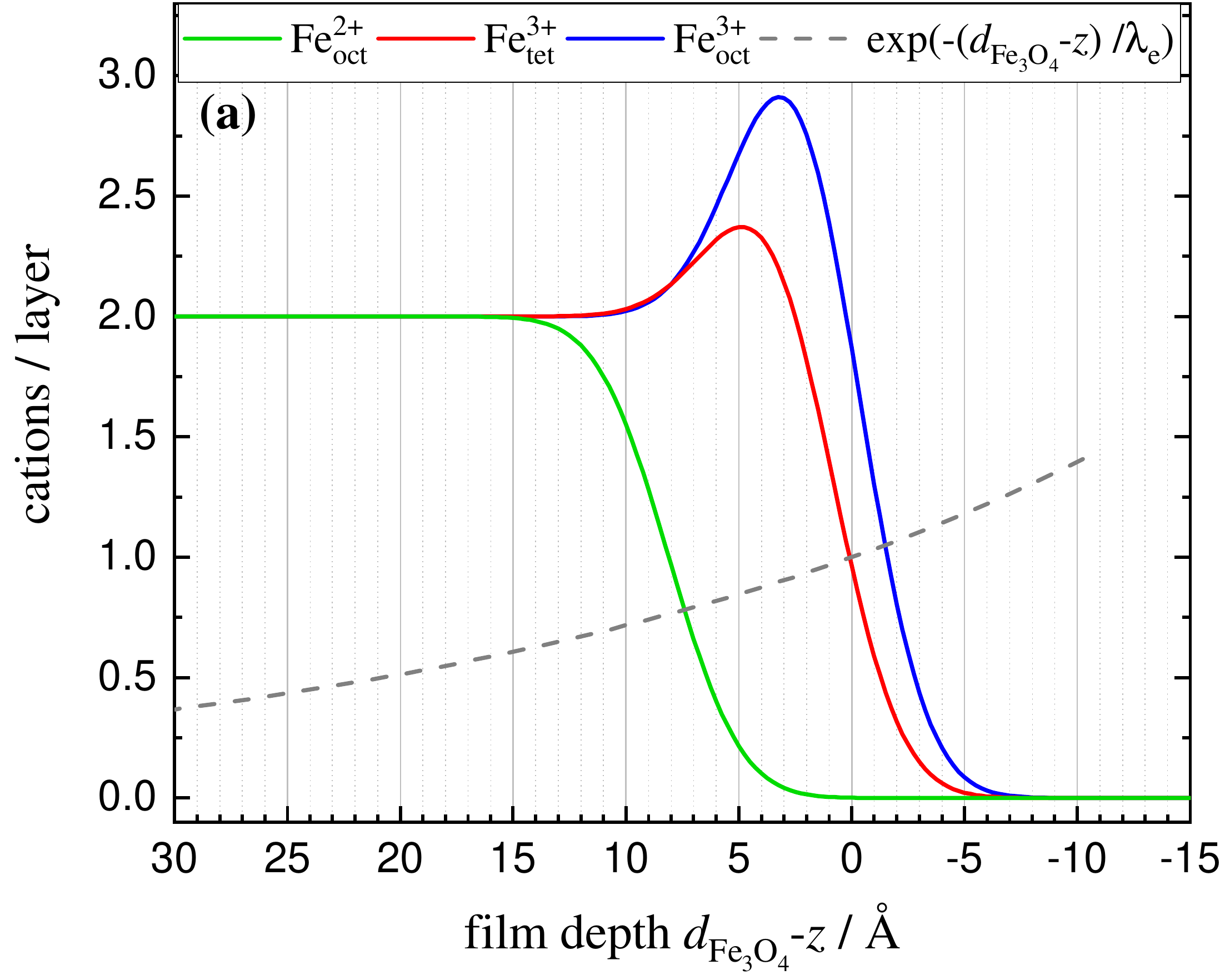}
}\hfill
\subfloat{
\includegraphics[scale=0.38]{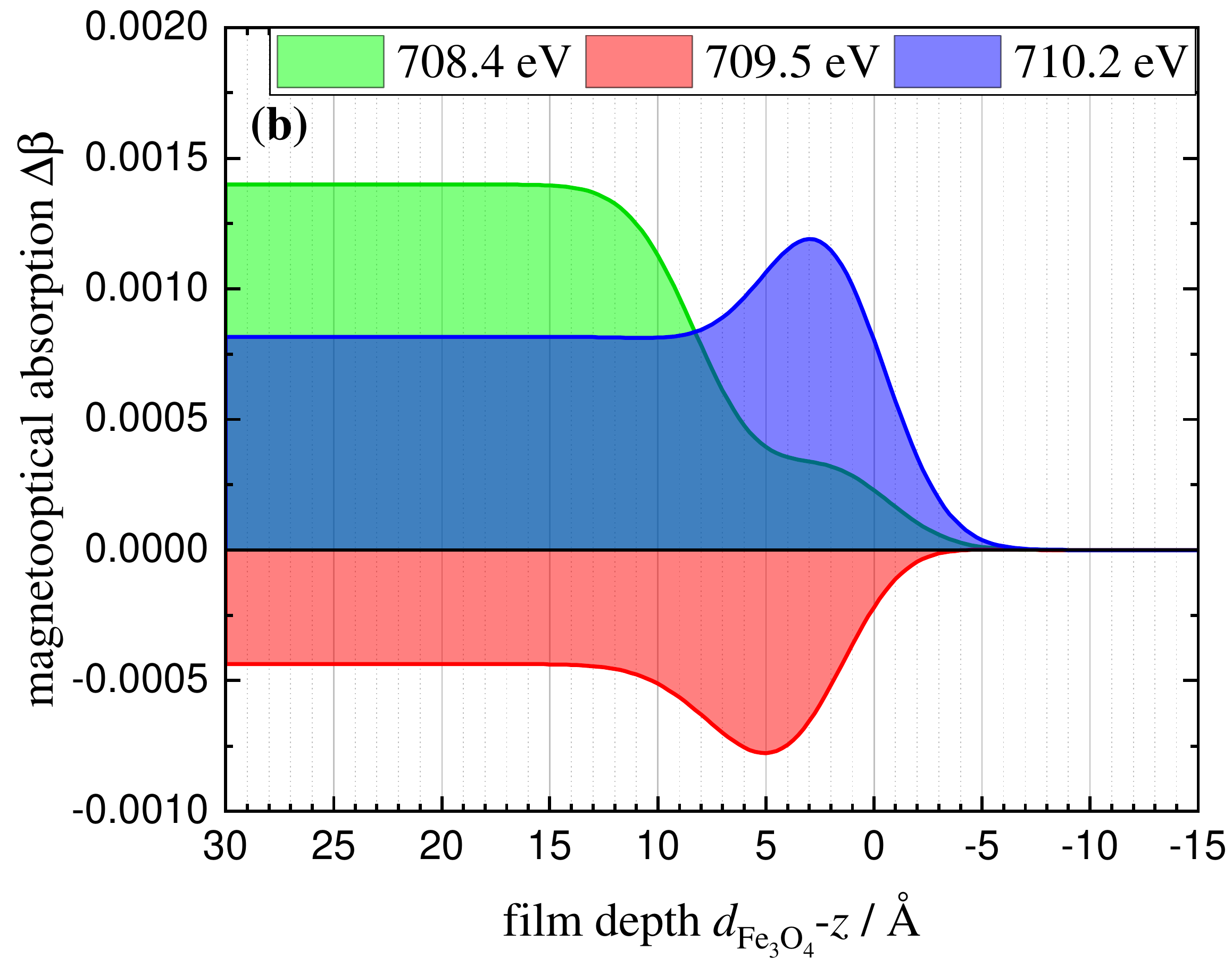}
}\\
\subfloat{
\includegraphics[scale=0.38]{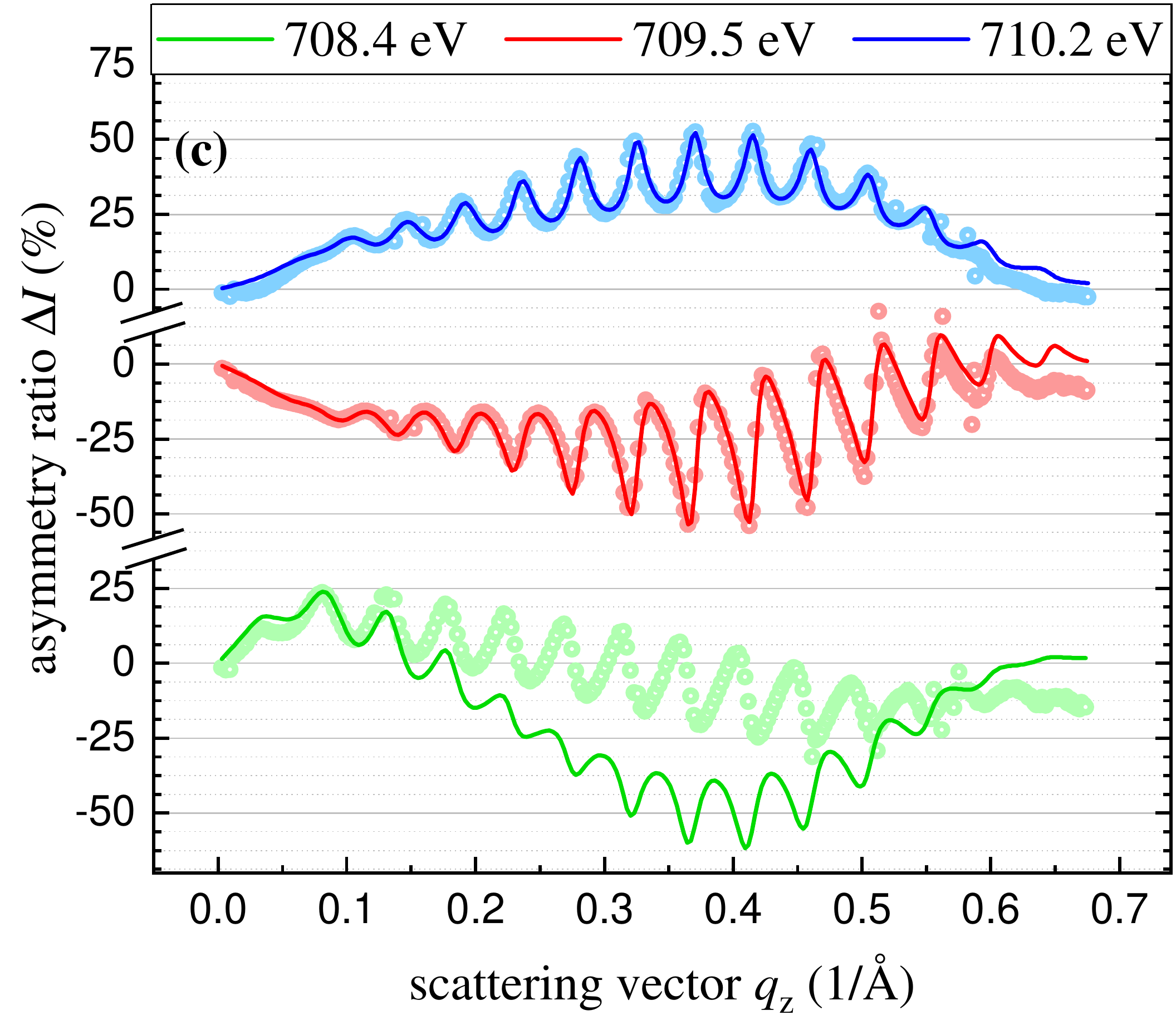}
}\hfill
\subfloat{
\includegraphics[scale=0.38]{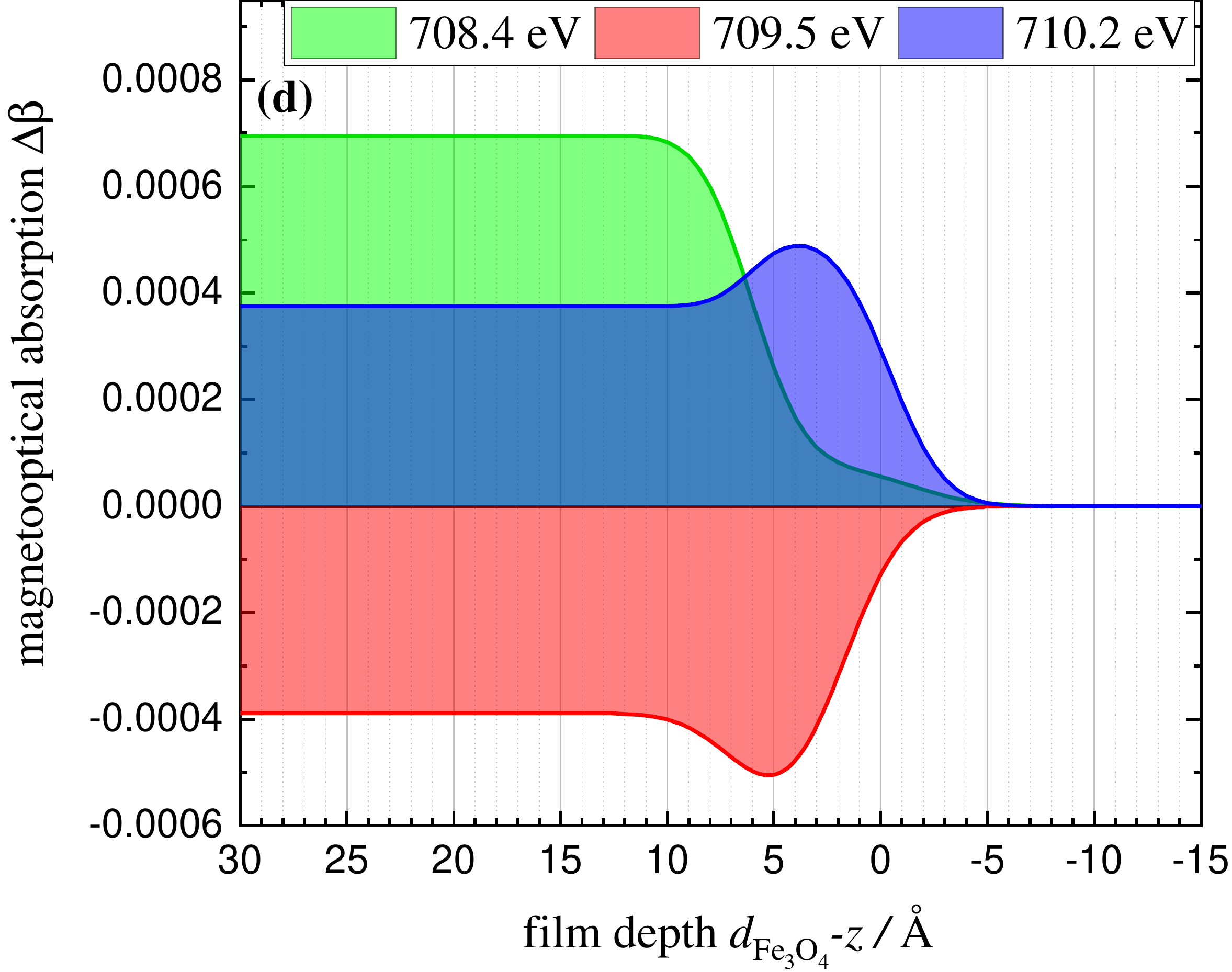}
}
\caption{(a) Cation depth profiles expected for a $\mathrm{Fe_3O_4}(001)$ film with a SCV surface and a roughness of $\sigma ^\mathrm{surf} = 2.5\,\mathrm{\r{A}}$. The dashed grey line shows the sensitivity of the TEY signal to electrons from film depth $d_\mathrm{Fe_3O_4}-z$, $\mathrm{exp}(-(d_\mathrm{Fe_3O_4}-z)/\lambda_\mathrm{e})$. (b) Expected magnetooptical depth profiles of the model film in (a) at the XMCD resonances $708.4\,$eV, $709.5\,$eV and $710.2\,$eV. (c) XRMR data from the $13\,$nm $\mathrm{Fe_3O_4}$ film, together with fits produced by the magnetooptical depth profiles in (d). (d) Magnetooptical depth profiles which are meant to recreate the profiles in (b), but slightly adapted to provide fits to the data in (c). For the resonances at $709.5\,$eV and $710.2\,$eV, satisfactory results can be achieved by this model, but not for the resonance at $708.4\,$eV.}
\label{SCV-Profiles}
\end{figure}
The magnetooptical depth profiles we find share some key features with the SCV structure of the $\mathrm{Fe_3O_4(001)}$ surface \cite{Bliem14}, but are not fully identical with it.
We want to emphasize that our samples had been exposed to air when the XRMR measurements were done, so a fully intact SCV surface is likely not present. In this chapter we present a thorough comparison of the SCV model to our findings. 
Along the [001] direction, the bulk structure of $\mathrm{Fe_3O_4}$ 
consists of alternating layers containing $(2~\mathrm{Fe^{2+}_{oct}}+2~\mathrm{Fe^{3+}_{oct})}$ per unit cell (u.c.) and $2~\mathrm{Fe^{3+}_{tet}/u.c.}$. 
DFT calculations of the SCV surface predict a different cation distribution in the first 4 cation layers:
The first cation layer consists of $4~\mathrm{Fe^{3+}_{oct}/u.c.}$, the second layer of $3~\mathrm{Fe^{3+}_{tet}/u.c.}$, the third layer of $2~\mathrm{Fe^{3+}_{oct}/u.c.}$, and the fourth layer of $2~\mathrm{Fe^{3+}_{tet}/u.c.}$. From there, $\mathrm{Fe_3O_4}$ continues with the cation distribution well-known from the bulk structure.

Figure \ref{SCV-Profiles}(a) shows the cation profiles as expected from an ideal SCV-surface with a surface roughness  $\sigma ^\mathrm{surf} = \mathrm{2.5\,\r{A}}$, corresponding to our films. Using the contributions of the cations to the XMCD spectrum in Tab. II of the main text, the expected magnetooptical depth profiles at the three XMCD resonances can be constructed, and are displayed in Fig. \ref{SCV-Profiles}(b). We recreated models of these theoretical magnetooptical depth profiles, and fitted their resulting asymmetry ratios $\Delta I$ to the XRMR data of the $13\,$nm $\mathrm{Fe_3O_4}$ film by retaining the main features of the SCV model. The XRMR data and fits are plotted in Fig. \ref{SCV-Profiles}(c), and the magnetooptical depth profiles which produce them in Fig. \ref{SCV-Profiles}(d).
The fits for the resonances at $\mathrm{709.5\,eV}$ and $\mathrm{710.2\,eV}$ describe the data very well, but not at $\mathrm{708.4\,eV}$. We can conclude that our data are consistent with the $\mathrm{Fe^{3+}}$ enrichment predicted by the SCV model, but not with the lack of $\mathrm{Fe^{2+}}$. 

\begin{figure}[t]
\centering
\includegraphics[scale=0.64]{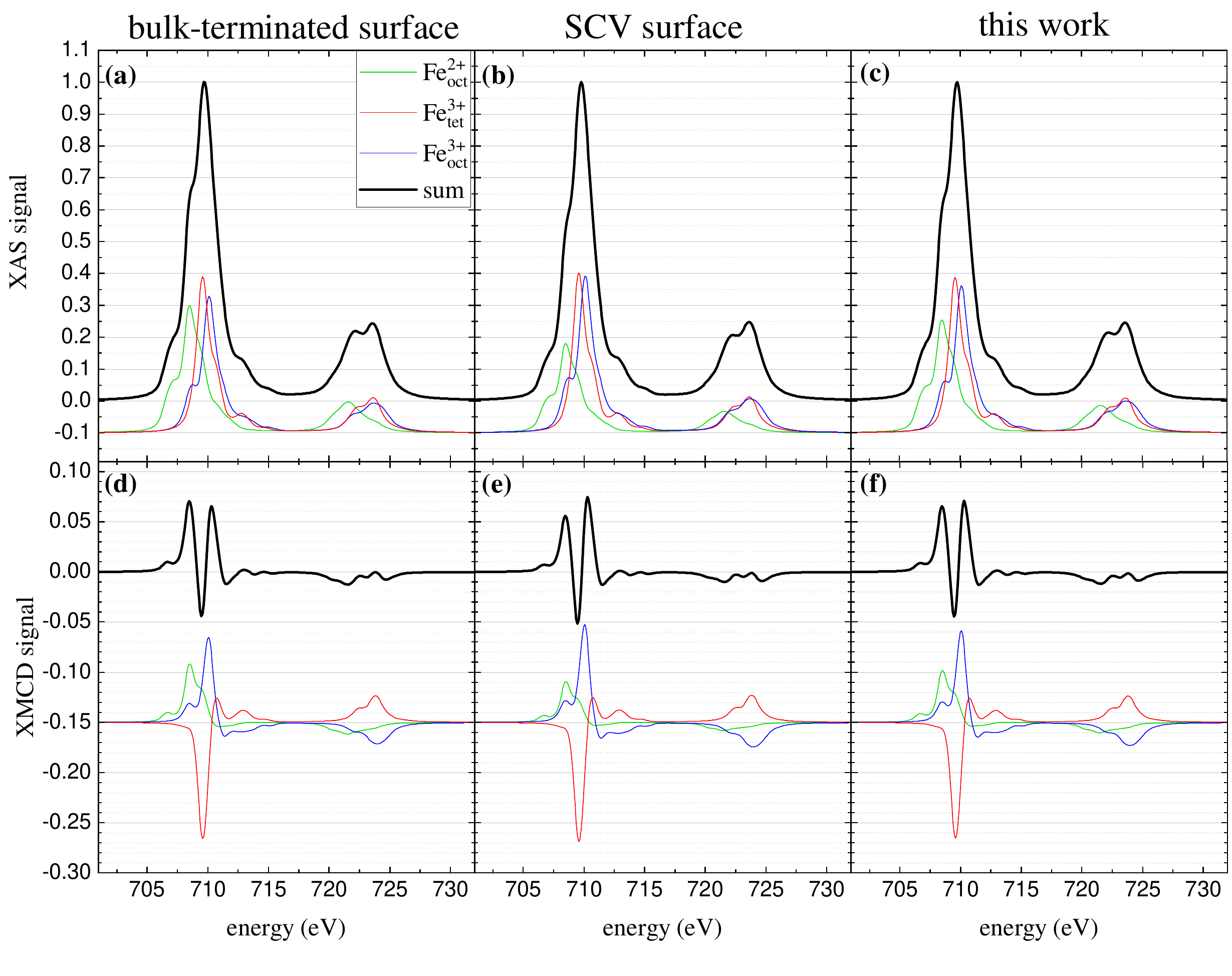}
\caption{Multiplet calculations of the XAS and XMCD spectra from (a),(d) a surface with bulk cation distribution, (b),(e) a SCV surface and (c),(f) the surface model we extract from the XRMR measurements in this work. They were calculated by weighting the individual cation spectra using Eq. \ref{eqWgt} with corresponding cation depth profiles $c_\mathrm{cation}(\zeta)$.}
\label{SCV-XAS}
\end{figure}

An $\mathrm{Fe^{3+}}$ enrichment of the $\mathrm{Fe_3O_4}(001)$ surface has some consequences for surface-sensitive techniques which are often utilized for the study of magnetite. We want to discuss this effect for the XAS and XMCD measurements used in this work. Figure \ref{SCV-XAS} shows multiplet simulations of the XAS and XMCD spectra of $\mathrm{Fe_3O_4}$ for three different surface compositions, assuming a TEY detection scheme. In TEY mode, the probing depth is limited by the electron escape depth $\lambda _\mathrm{e}$, which is about $30\,\mathrm{\r{A}}$ for $\mathrm{Fe_3O_4}$. In Fig. \ref{SCV-Profiles}(a) the cation depth profiles $c_\mathrm{Fe^{2+}_{oct}}(\zeta)$,$c_\mathrm{Fe^{3+}_{tet}}(\zeta)$, $c_\mathrm{Fe^{3+}_{oct}}(\zeta)$ for films with SCV surface are shown together with the exponential decay $\mathrm{exp}(-\zeta /\lambda_\mathrm{e})$ describing the contribution to the TEY signal from film depth $\zeta = d_\mathrm{Fe_3O_4}-z$. Integration over their product yields a weight factor
\begin{equation}
wgt_\mathrm{cation} = \int c_\mathrm{cation}(\zeta) \cdot \mathrm{exp}(-\zeta/\lambda_\mathrm{e})\mathrm{d\zeta},
\label{eqWgt}
\end{equation}
which we used to weight the individual spectra of the cations in Fig. \ref{SCV-XAS}.

Figure \ref{SCV-XAS}(a) and \ref{SCV-XAS}(d) show the XAS and XMCD spectra for a bulk-terminated surface with a 1:1:1 ratio of the three cations, \ref{SCV-XAS}(b) and \ref{SCV-XAS}(e) for an SCV surface as illustrated in \ref{SCV-Profiles}(a), and \ref{SCV-XAS}(c) and \ref{SCV-XAS}(f) for the model extracted from the XRMR data in this work. For the multiplet simulations, we used the same parameters as for the fits to the XMCD spectra of the $25\,$nm $\mathrm{Fe_3O_4}$ film recorded at BESSY. They can be found in Tab. \ref{tab:mp}.

The XAS spectra differ only very slightly because of the strong overlap of the three individual spectra, and only the low-energy shoulder at the $\mathrm{Fe^{2+}_{oct}}$ becomes less pronounced for the $\mathrm{Fe^{3+}}$-rich surfaces. In the XMCD spectra, the change is more easily visible in the intensity difference of the resonances at $708.4\,\mathrm{eV}$ and $710.2\,\mathrm{eV}$. This difference can lead to a slight underestimation of $\mathrm{Fe^{2+}_{oct}}$ in a multiplet analysis. In fact, our XMCD spectra compare best with the model in Fig. \ref{SCV-XAS}(f), but the difference observed between Fig. \ref{SCV-XAS}(f) and Fig. \ref{SCV-XAS}(d) is well in the range of common sample-to-sample variation.

For the sum rule analysis, the impact can be expected to be small, since the underestimation of $\mathrm{Fe^{2+}_{oct}}$ is compensated by an overestimation of $\mathrm{Fe^{3+}_{oct}}$ of almost same extent in the XMCD signal.

\bibliography{CationProfilesBib}